\documentclass[times]{TRR}

\usepackage{moreverb,url}

\usepackage{algorithm}
\usepackage{algpseudocode}

\usepackage[acronym]{glossaries}
\usepackage{float}
\usepackage{graphicx}

\usepackage{xcolor}

\newacronym{indot}{INDOT}{Indiana Department of Transportation}

\usepackage{hyperref}

\newcommand\BibTeX{{\rmfamily B\kern-.05em \textsc{i\kern-.025em b}\kern-.08em
T\kern-.1667em\lower.7ex\hbox{E}\kern-.125emX}}

\begin{document}

\runninghead{Naman, Zhang, Ault and Krogmeier}

\title{Automating Work Orders and Tracking Winter Snow Plows and Patrol Vehicles with Telematics Data}

\author{Anugunj Naman, Aaron Ault, Yaguang Zhang and James Krogmeier}

\affiliation{Affliliation: Purdue University, West Lafayette, Indiana, USA}

\corrauth{Anugunj Naman, anaman@purdue.edu}

\begin{abstract}
Winter road maintenance is a critical priority for the \acrlong{indot}, which manages an extensive fleet across thousands of lane miles. The current manual tracking of snowplow workloads is inefficient and prone to errors. To address these challenges, we developed an in-browser web application that automates the creation and verification of work orders using a large-scale GPS dataset from telematics systems. The application processes millions of GPS data points from hundreds of vehicles over winter, significantly reducing manual labor and minimizing errors. Key features include geohashing for efficient road segment identification, detailed segment-level work records, and robust visualization of vehicle movements, even on repeated routes. Our proposed solution has the potential to enhance the accuracy and granularity of work records, support more effective resource allocation, ensure timely compensation for drivers, alleviate administrative burdens, and allow managers to focus on strategic planning and real-time challenges. The web application can be accessed at \href{https://github.com/oats-center/arrtrack/}{https://github.com/oats-center/arrtrack/}

\hfill\break%
\end{abstract}

\maketitle

\newpage

\section{Introduction}
In the United States, over 70\% of roads are found in regions prone to snow, similar to conditions in Indiana~\cite{FederalHighwayAdministration2023}. Snow and ice on roads not only slow down traffic but also pose serious safety risks for drivers~\cite{FederalHighwayAdministration2023}. Due to this, the \gls{indot} places snow and ice removal during the winter months as its number one priority~\cite{Zhang2024}. This massive effort involves more than 1000 snow plows, nearly 2000 employees, and over 29,000 lane miles of roads~\cite{IndianaDepartmentOfTransportation}. Managing such an extensive fleet is challenging, particularly because tracking and reporting the workloads of all these vehicles is done manually~\cite{ IndianaDepartmentOfTransportation2023}. Work records for these vehicles are currently created by hand, which can be time-consuming and limited in recording granularity, especially given the urgency and weather-dependent nature of winter operations~\cite{ IndianaDepartmentOfTransportation2023}. 

To address these challenges, it is essential to consider the significant administrative burden faced by managers during winter operations. A manager responsible for overseeing numerous vehicles may have to generate hundreds or even thousands of work order entries each week during snowy periods. These records are crucial as they directly impact driver payroll, the procurement of treatment materials, and the planning of future road improvements. During storm events, drivers often work long hours, making the manual logging of work orders not only a distraction but also a significant hassle. The burden of this manual process is substantial. In severe snowstorms, the sheer volume of work orders can overwhelm the manual approach, potentially causing significant delays in processing drivers' salaries. This delay can lead to dissatisfaction among drivers who rely on timely payment after their extended hours of hard work. Moreover, from the managers' perspective, creating these work records is a necessary but low-value task. Their time and attention are better spent on higher-value activities, such as strategic planning, resource allocation, and responding to real-time challenges that arise during snowstorms. Hence, automating the work order process would not only alleviate the administrative burden but also allow managers to focus on these more critical and impactful responsibilities.

Research has been heavily focused on automating logistics and decision-making using telematics based on satellite navigation technologies~\cite{Ghaffarpasand2022}. This includes recognizing human activities~\cite{Wang2021}, monitoring agricultural vehicles~\cite{Zhang2020}, and planning routes for self-driving cars~\cite{Kinable2016}. However, most of the current research is limited to small-scale, short-term studies rather than extensive, long-term analyses, such as performance evaluation for fleet management~\cite{Goel2008}. Collecting and making use of telematics data for an entire fleet over a large area, especially in rural regions like much of Indiana, is particularly difficult. As a result, studies often focus on a few vehicles or objects of interest~\cite{Zhang2020,Goel2008}, lacking comprehensive trend and pattern analysis over time. Winter road maintenance presents unique challenges, as vehicles often travel the same routes multiple times, unlike the single-trip cases typically studied~\cite{Zhang2020,Kinable2016}. Visualizing these repeated movements is complex.  

Telematics data analysis can be highly application-specific, requiring tailored algorithms for optimal results~\cite{Zhang2017}, even for similar tasks within \gls{indot}'s responsibilities. A complicating factor in GPS-centric fleet telematics analysis is the use of a linear reference system with road names and mileposts instead of GPS coordinates within \gls{indot}~\cite{Zhang2024}.  Converting GPS coordinates from telematics trackers to the road segments of the \gls{indot} reference system is non-trivial because it involves not just a simple translation but also aligning these coordinates accurately to the corresponding road segments and mileposts in \gls{indot}'s database. This requires sophisticated mapping algorithms and precise calibration to ensure that the GPS data aligns correctly with the linear reference system, accounting for various anomalies and discrepancies such as GPS drift, road curvature, and varying segment lengths.

Our work focuses on using a large-scale, fleet-level GPS dataset for \gls{indot} vehicles to improve the work recording process, both in time and quality, compared to current methods. Sourced from the \gls{indot} Data Warehouse, the GPS dataset encompasses 5,115,844 point entries, representing the winter maintenance operations of 1051 vehicles throughout a complete winter season (12/2/2020--4/30/2021). It covers 50,829 tracks across the entire state of Indiana. The sampling time statistics reveal that the smallest interval between adjacent points is 1 minute. 34\% of the intervals are exactly 1 minute, while 14\% extend beyond 5 minutes. This extensive dataset, with its multitude of record points, poses challenges for manual inspection and analysis~\cite{Zhang2024}. Our proposed solution is an in-browser web application that reads GPS tracks generated by \gls{indot}'s telematics provider, verifies existing work records against the known tracks of the vehicles, and can create new work records with start and end time tied to specific road segments. We summarize the contributions of our web application as follows:

\begin{enumerate}
    \item \textbf{Less Human Labor}:
    \begin{itemize}
        \item With the app, the tedious process of manually recording work activity has been automated, significantly reducing the need for human intervention and minimizing errors.
    \end{itemize}

    \item \textbf{Improved Accuracy and Detail in Work Records}:
    \begin{itemize}
        \item Work records created by the app offer precise road-segment-level granularity, capturing detailed data for each segment with consistent logging of timestamps. This ensures that even minute details are accurately recorded.
        \item The app allows for the visualization of detailed paths over time, even when snow plows travel back and forth over the same road, ensuring comprehensive tracking of all movements.
    \end{itemize}

    \item \textbf{Enhanced Cost Analysis and Planning}:
    \begin{itemize}
        \item The segment-based path tracking offered by the app is essential for accurate cost analysis of each road segment. This capability aids in precise cost evaluation and supports future planning efforts.
    \end{itemize}

    \item \textbf{User-Friendly and Secure Application}:
    \begin{itemize}
        \item The browser-based application works locally, ensuring privacy since no data needs to be uploaded to the cloud. Its web-based nature, hosted on GitHub Pages, eliminates the need for a dedicated server, making it easily accessible to anyone.
        \item Despite being limited to consumer-grade hardware in the browser, the app has been optimized to handle millions of GPS points from thousands of fleet vehicles, showcasing its efficiency and robustness.
    \end{itemize}

    \item \textbf{Bridging Research and Practical Use}:
    \begin{itemize}
        \item This application effectively bridges the gap between university research and practical use for \gls{indot} and similar organizations. It makes advanced research analysis accessible for general purposes by eliminating the need for direct system integration with IT infrastructure and the need for long-term server maintenance, enhancing the practical utility of academic advancements.
    \end{itemize}
\end{enumerate}

We thus aim to automate the process of manual recording of work orders for winter road maintenance operations, making them more efficient and accurate. 

\section{Proposed System}

In this section, we present our web application designed to automate work orders and manage fleet operations. We begin with a comprehensive description of the web application’s features and functionalities. Following this, we present the detailed algorithms used to calculate the snow operation time for a specific vehicle on a given day and specific road segments.

\subsection{Web Application}

\begin{figure}[h]
    \includegraphics[width=0.95\linewidth]{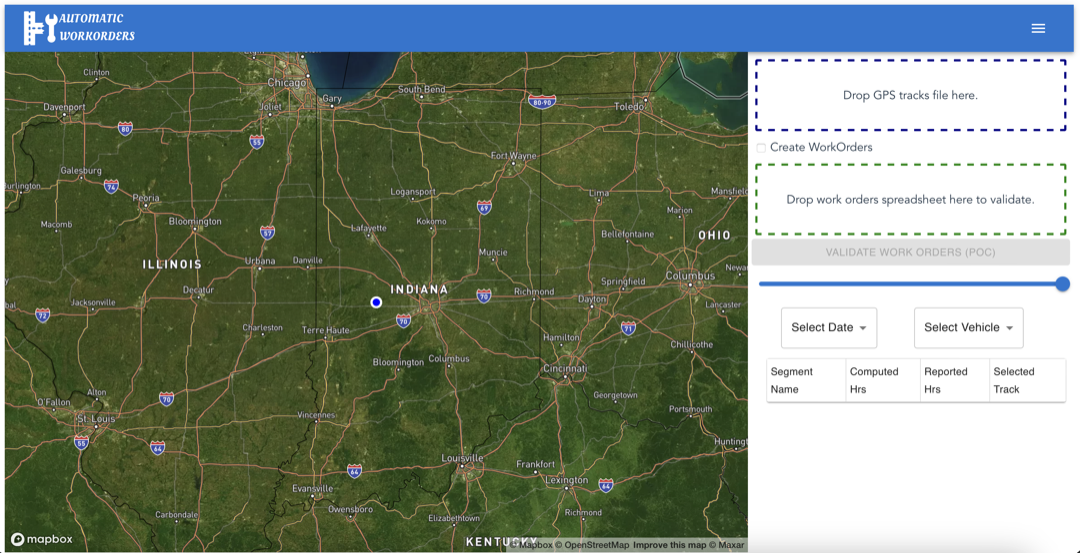}
    \caption{Initial view of the web application.}
    \label{initial_interface}
\end{figure}

The initial interface of the web application, as depicted in \textbf{Figure \ref{initial_interface}}, is systematically divided into two primary sections: a map and a control panel. This dual-section design works well in allowing simultaneous access to both controls and map visualization on a laptop or desktop-sized, landscape-oriented screen which is the targeted use case.

\begin{figure}[h]
\includegraphics[width=0.95\linewidth]{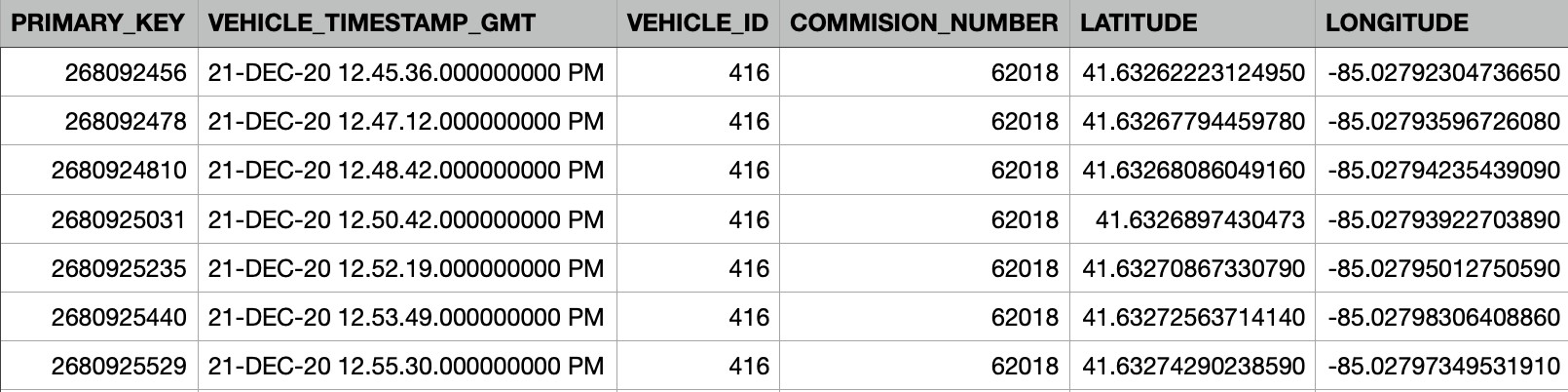}
\caption{Sample GPS data.}
\label{sample_gps_sheet}
\end{figure}

The map interface, powered by Mapbox, occupies the left and central portions of the screen. This dynamic and interactive feature provides a satellite view of Indiana, highlighting all major state highways and interstates, and delineating state boundaries. This helps users, particularly fleet managers and operators, to visualize their vehicles' locations and movements across the region. A prominent blue marker is displayed at the start location of a selected vehicle, allowing users to quickly identify the vehicle's initial position. This marker represents the vehicle's "current location" corresponding to the position of the time slider in the control panel, while a line element is drawn to represent the vehicle's historical path from the start of the day up to that synthetic "current" location. The line segment feature, which visually represents the vehicle's route on the chosen day, will be described and shown in detail later.

\begin{figure}[h]
\includegraphics[width=0.95\linewidth]{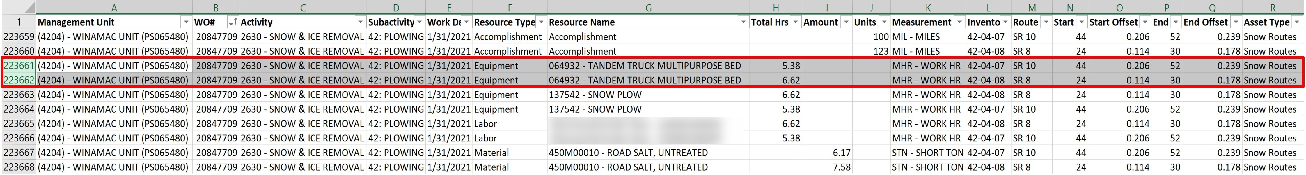}
\caption{Sample work order records. Note that multiple entries for the same vehicle on the same day may appear, as highlighted by the red rectangle.}
\label{sample_work_order_sheet}
\end{figure}

\begin{figure}[h]
    \includegraphics[width=0.95\linewidth]{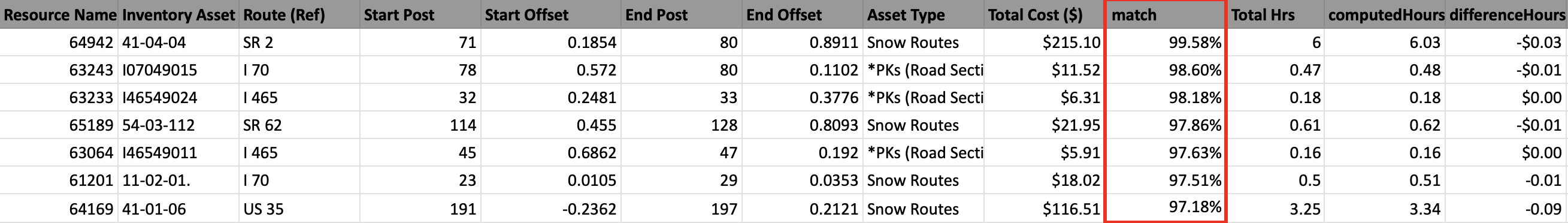}
    \caption{Sample verification report. The "Match" column, highlighted by the red rectangle, indicates how closely the computed time aligns with the reported time.}
    \label{sample_verify_report}
\end{figure}

To the right of the map is the control panel, divided into several interactive sections designed to enhance the management and tracking of vehicle activities. The upper segment as seen in \textbf{Figure \ref{initial_interface}} of this panel features two prominent file upload areas. The top box, outlined with a blue dashed border, is designated for uploading GPS track files. This feature allows users to input data directly from their fleet's GPS systems, ensuring accurate tracking and analysis of vehicle movements.  Below this, a green dashed box is reserved for work order spreadsheets. \textbf{Figure \ref{sample_gps_sheet}} and \textbf{Figure \ref{sample_work_order_sheet}} show examples of the GPS data and work order records that are used as input for the application, respectively. Upon uploading a valid work order spreadsheet, the \textit{Validate Work Orders} button becomes active, initiating the validation process. After validation, you can download the validation report. An example of a validation report can be seen in \textbf{Figure \ref{sample_verify_report}}.

\begin{figure}[h]
    \includegraphics[width=0.95\linewidth]{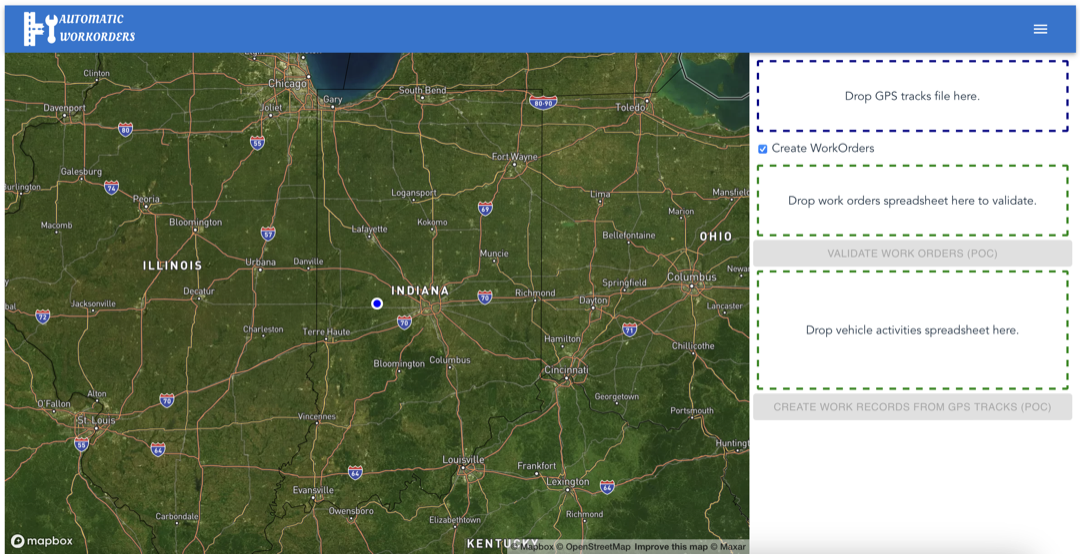}
    \caption{Work order creation interface.}
    \label{create_orders_interface}
\end{figure}

\begin{figure}[h]
    \includegraphics[width=0.95\linewidth]{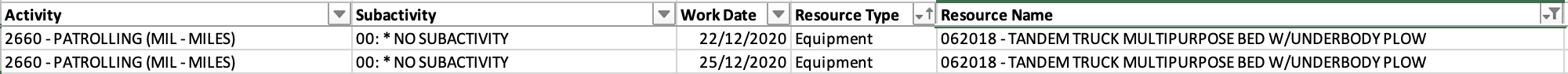}
    \caption{Sample vehicle activity data.}
    \label{sample_vehicle_activity}
\end{figure}

\begin{figure}[h]
    \includegraphics[width=0.95\linewidth]{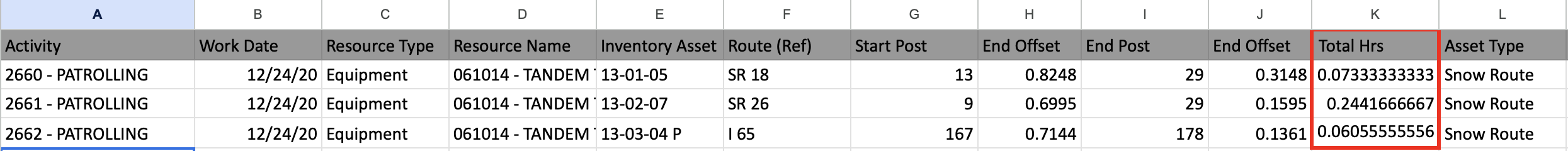}
    \caption{Sample creation report. The "Total Hrs" column, highlighted by the red rectangle, indicates the computed time for the given road segment.}
    \label{sample_created_records}
\end{figure}

In the control panel, a checkbox labeled \textit{Create Work Orders} allows users to initiate the creation of work orders directly from the interface by uploading a list of vehicle activities, thereby automating the tedious process of manually recording work activity for each covered road segment. The work order creation interface is shown in \textbf{Figure \ref{create_orders_interface}}. An example of vehicle activities used as input for work order creation can be seen in \textbf{Figure \ref{sample_vehicle_activity}}. An example of created work orders can be seen in \textbf{Figure \ref{sample_created_records}}. 

This application also extends its functionality through the inclusion of supplementary tools designed for targeted data selection and manipulation. These tools, as seen in the middle of the right panel in \textbf{Figure \ref{initial_interface}}, are in the form of drop-down menus labeled \textit{Select Date} and \textit{Select Vehicle}. They allow users to implement granular filters, enabling the visualization and analysis of data about specific dates or individual vehicles within the fleet. The tools become available when the checkbox \textit{Create Work Orders} is disabled, allowing users to customize their interaction with the data as needed for work order verification. This feature proves particularly advantageous in facilitating historical data analysis and the focused tracking of specific vehicles. Once a desired vehicle ID and date are selected, the application leverages GPS data to generate a visual representation of the chosen vehicle's track.

\begin{figure}[h]
    \includegraphics[width=0.95\linewidth]{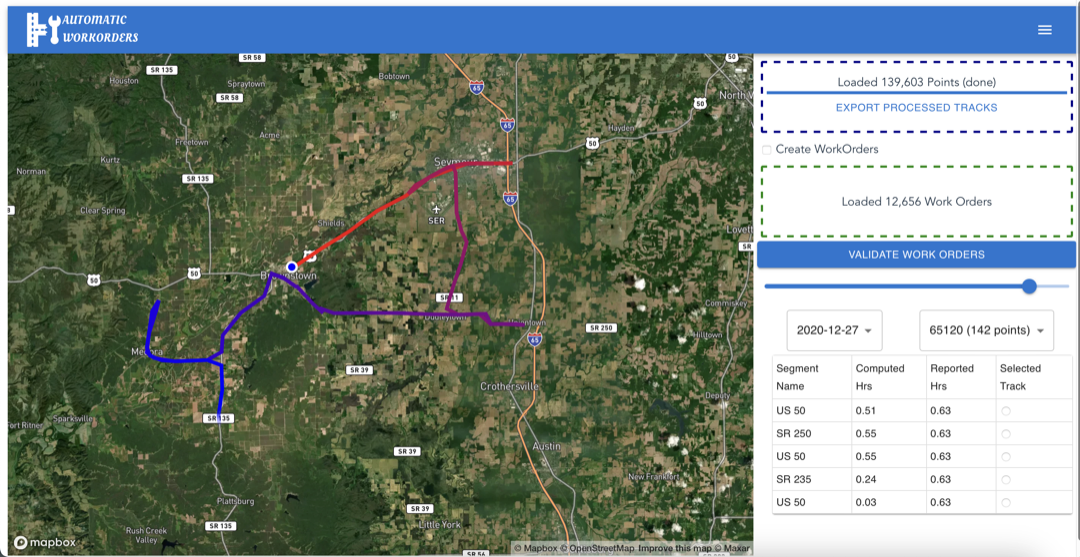}
    \caption{Interface displaying a sample case with the vehicle's path traced from GPS data.}
    \label{sample_case_normal}
\end{figure}

Further customization options are facilitated by a horizontal slider positioned beneath the validation button. This slider as seen in \textbf{Figure \ref{initial_interface}} allows users to dynamically adjust the displayed timeframe, enabling the tracking of a particular vehicle's path through the use of animated lines. These lines are color-coded by time intervals, with red signifying past GPS data points and blue highlighting the most recently traversed location. This implementation injects a dynamic element into the application, empowering users to gain valuable insights into vehicle movements across the temporal dimension. A visual representation of a vehicle's tracked path can be found in \textbf{Figure \ref{sample_case_normal}}.

At the bottom of the control panel, there is a table as seen in \textbf{Figure \ref{sample_case_normal}} designed to show detailed information about the chosen vehicle's operations during work order verification. The table has columns labeled \textit{Segment Name}, \textit{Computed Hrs}, \textit{Reported Hrs}, and \textit{Selected Track}. These columns provide specific insights into the fleet's activities. \textit{Segment Name} shows the specific road segment the vehicle traveled on that day. \textit{Computed Hrs} and \textit{Reported Hrs} provide a comparison between the exact time spent on each road segment, as calculated from the GPS data timestamps, and the time reported for those segments in the system records. This allows for accurate verification of reported times and identification of any discrepancies.

\begin{figure}[h]
    \includegraphics[width=0.95\linewidth]{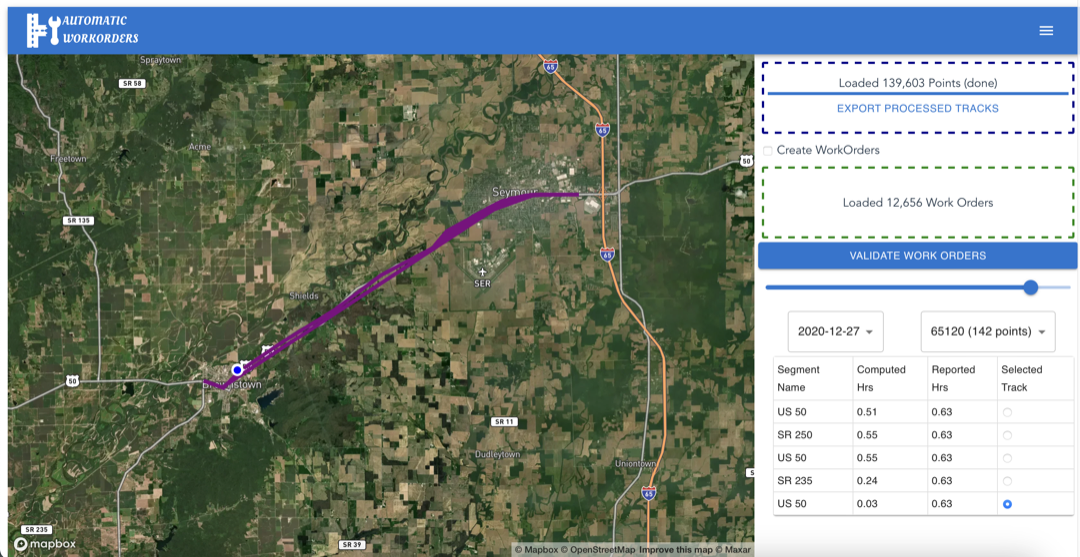}
    \caption{Interface with a working example case illustrating the specific road segment path traced by the vehicle using GPS data.}
    \label{sample_segment_case}
\end{figure}

The \textit{Selected Track} column contains radio buttons allowing users to focus on specific road segments rather than the entire route. By selecting a radio button, users can isolate and examine the details of a particular segment. This table provides a clear summary of the selected vehicle's activities, enabling managers to quickly assess performance, identify discrepancies, and make informed decisions to improve fleet operations. A working visual example of a specific segment the vehicle traveled, along with the table summarizing the fleet's activities for each road segment on the selected date, can be seen in \textbf{Figure \ref{sample_segment_case}}.

Overall, the application design allows for robust visualization of detailed paths over time, even when snow plows travel back and forth over the same road.

\subsection{Algorithm Formulation}

\begin{figure}[h]
    \includegraphics[width=0.95\linewidth]{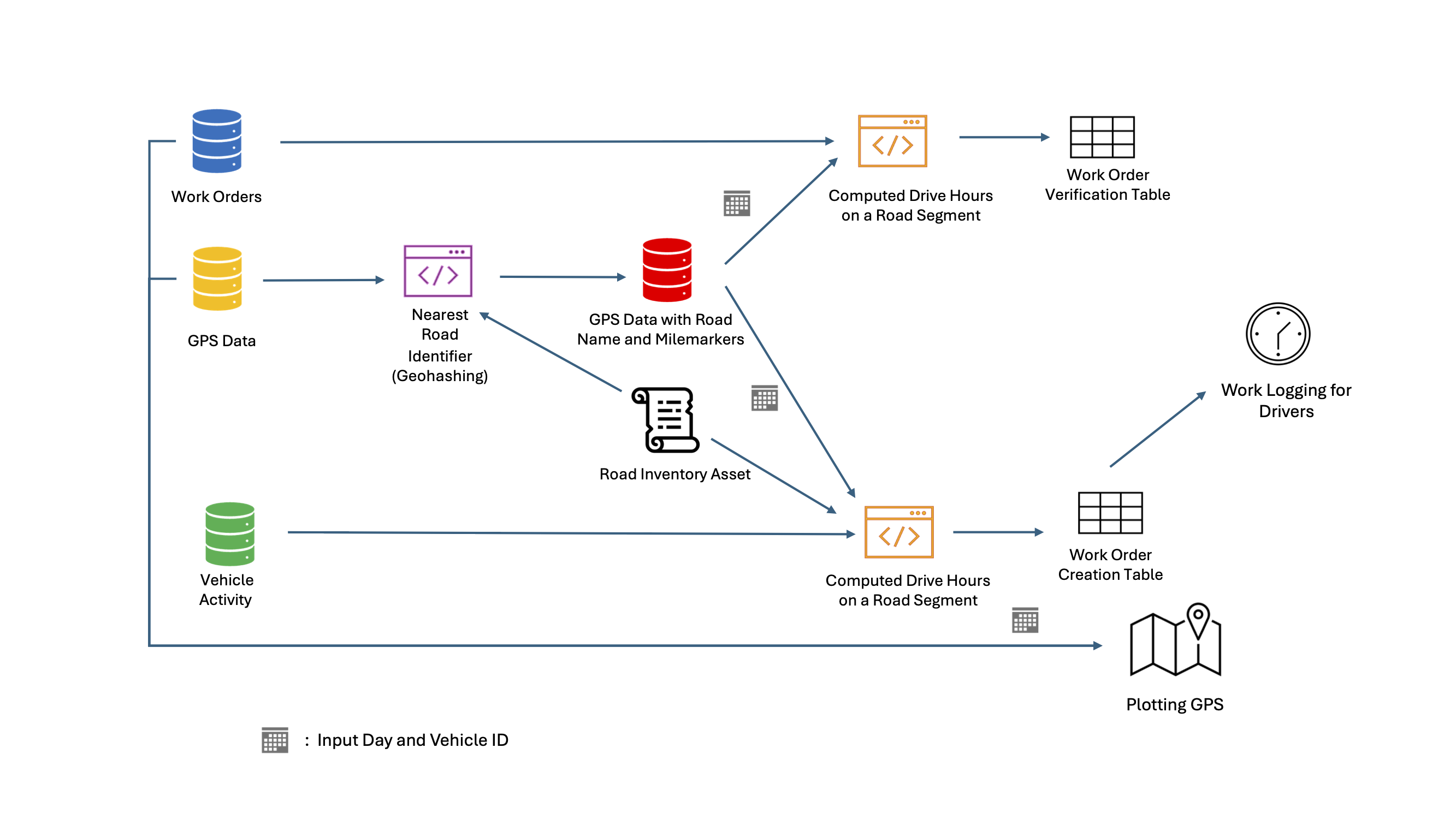}
    \caption{An overview of the application, demonstrating how each input is processed by the backend algorithms to generate and display results in the web application.}
    \label{overview}
\end{figure}

In this section, we explain the two algorithms we developed to analyze telematics data for our web application, which are used to calculate the \textit{Computed Hours} in the summary table visible in the web application. \textbf{Figure \ref{overview}} provides a visual summary of the backend process, illustrating how these algorithms interact to produce the metrics displayed in the web application.

The first part of the algorithm determines the most likely road the vehicle is on at each GPS point or decides if the vehicle is off the road. To handle the large number of points in the INDOT dataset efficiently, the algorithm uses geohashing~\cite{Niemeyer2008} to group nearby road segments together and then find the relevant group for each point. This process is illustrated as the pink program in Figure \ref{overview}. After identifying the roads and mile marker offsets for each GPS point, the second part of the algorithm calculates the \textit{Computed Hours}, which represents the total time the vehicle was operating on each road segment that day. This step is shown as the orange program in \textbf{Figure \ref{overview}}.

Before diving into the algorithm, we also define a few key terms that will be referenced in the following sections:

\begin{itemize}
    \item \textbf{Route Ref}: This is the road name or reference identifier for the road segment.
    \item \textbf{Road Type}: This refers to the classification of the road, such as interstate, highway, state highway, etc.
    \item \textbf{Mile Markers}: These are numerical values representing specific locations along a road, typically indicating distance from a defined starting point.
    \item \textbf{Start Post and End Post}: These are the mile markers that denote the beginning and end of a road segment.
    \item \textbf{Road Inventory Assets}: These are the comprehensive collection of road segments used by \gls{indot} to facilitate maintenance, planning, and operational efficiency.
\end{itemize}

\subsubsection{Geohashing for Nearest Road Identification}

For calculating \textit{Computed Hours} for each road segment the vehicle traveled on that day, we first need to ensure that the current GPS point is on the mentioned road segment. Identifying the closest road to a GPS point is a complex problem due to the large number of roads and GPS points involved. Each GPS point must be compared to every road, requiring a vast number of distance calculations. This brute-force approach is slow and inefficient, as it involves looping through all roads for every single GPS point, resulting in a huge computational load. Additionally, handling the full road dataset in one go is impractical, especially in memory-limited environments like web browsers, because the data is too large to download and process efficiently.

Geohashing offers an elegant solution to this problem. It encodes geographic coordinates into short strings of letters and numbers, each representing a small rectangular area on a map, called a tile. By dividing the entire map into these smaller tiles, geohashing creates a spatial index that makes it much easier to manage and search the data~\cite{Niemeyer2008}. \textbf{Algorithm \ref{ag2}} illustrates the pseudo-code implementation. When a GPS point needs to be processed, its geohash is calculated to determine which tile it falls into. Only the roads within this tile and its neighboring tiles need to be considered, drastically reducing the number of distance calculations required (Algorithm~\ref{ag2}, lines 3-5).

\begin{algorithm}[h]
\begin{algorithmic}[0]
\Function{IdentifyRoads}{GPSPoints, Roads}
    \State GeohashMap $\gets$ createGeohashMap(Roads)
    \For{each point in GPSPoints}
        \State currentGeohash $\gets$ computeGeohash(point)
        \State nearbyTiles $\gets$ getNearbyTiles(currentGeohash)
        \State candidateRoads $\gets$ retrieveRoads(GeohashMap, nearbyTiles)
        
        \State minDistance $\gets$ infinity
        \State closestRoad $\gets$ null
        \For{each road in candidateRoads}
            \State distance $\gets$ calculateDistance(point, road)
            \If{distance < minDistance}
                \State minDistance $\gets$ distance
                \State closestRoad $\gets$ road
            \EndIf
        \EndFor
        
        \State selectedRoad $\gets$ null
        \If{isHintRoadValid(previousRoad, point)}
            \State selectedRoad $\gets$ previousRoad
        \ElsIf{isCloseEnough(closestInterstate, point)}
            \State selectedRoad $\gets$ closestInterstate
        \ElsIf{isCloseEnough(closestStateRoad, point)}
            \State selectedRoad $\gets$ closestStateRoad
        \ElsIf{isCloseEnough(closestLocalRoad, point)}
            \State selectedRoad $\gets$ closestLocalRoad
        \Else
            \State selectedRoad $\gets$ OffRoad
        \EndIf
        
        \State updatePreviousRoad(selectedRoad)
    \EndFor
\EndFunction
\end{algorithmic}
\caption{Geohashing for Nearest Road Identification}
\label{ag2}
\end{algorithm}

This method not only speeds up the computation but also reduces the amount of data that needs to be downloaded to the browser at any one time. Instead of dealing with the entire road dataset, the algorithm can focus on a much smaller subset, making it feasible to run efficiently even in a web browser. This allows the spatially-indexed set of road segments to be hosted entirely in Github Pages as small, simple files named by their geohash, eliminating the need for any external database or server.  By limiting the search to a localized area, geohashing ensures that the process of finding the nearest road is both quick and resource-efficient (Algorithm~\ref{ag2}, lines 6-13).

The algorithm also employs a hierarchical decision-making process to select the most appropriate road. Initially, it checks if the previously identified road (what we call the \textit{hint road}) is still valid for the current GPS point (Algorithm~\ref{ag2}, line 16). If not, it prioritizes the nearest interstate roads, followed by state and local roads, based on their proximity (Algorithm~\ref{ag2}, lines 17-23). If no road is sufficiently close, the point is classified as off-road. This structured approach ensures that the road identification process is deterministic, resilient to accidental jumps when a vehicle crosses other roads, and it favors selecting interstates and state highways where INDOT snowplowing typically occurs. This geohashing and distance algorithm allows fast identification of roads for a given GPS data point significantly improving the performance and practicality of GPS-based road identification.

\subsubsection{Compute Seconds on Road Segment for Vehicle on Day}

In this section, we describe the second algorithm that calculates the number of hours a vehicle was actively engaged in snow plowing and patrolling operations on \textit{each specific road segment} listed in the work order records for a given day. This provides detailed data for the summary table, as shown in \textbf{Figure \ref{sample_segment_case}}.

The algorithm works by processing each GPS data point for a vehicle on the selected road segment from the work orders. It sums the valid driving durations between successive GPS points while ignoring periods when the vehicle is stationary or the data is unreliable. \textbf{Algorithm \ref{ag3}} presents the pseudo-code for this implementation.

The algorithm starts by taking the vehicle ID and day information along with a road segment as input and verifies the presence of a route reference, denoted as \texttt{Route Ref}, for the input segment \texttt{seg}. This route reference serves as the identifier for the specific road under consideration. If this reference is absent, it is impossible to determine the pertinent road segment which leads to a return value of 0 seconds. This precautionary step ensures that the algorithm only processes segments with identifiable route information, thereby maintaining data integrity and avoiding erroneous calculations.

We then utilize the \texttt{roadNameToType} function to discern the road type and name from the route reference. This function is pivotal as it standardizes the identification of road types, such as interstate highways or local roads. By initializing the \texttt{startpost} and \texttt{endpost} variables along with their respective offsets, the algorithm sets the stage for accurate segment delineation.

Mile markers are indispensable for precise localization along a road. They provide exact reference points, facilitating accurate distance and time measurements. If either the start post or end post is missing, the algorithm logs this discrepancy and returns 0 seconds. This ensures that calculations are only performed when both mile markers are available, thus preventing inaccuracies due to incomplete data. To maintain logical consistency, the algorithm assesses and adjusts the start and end posts based on the presence of positive or negative offsets. For instance, a negative start offset necessitates adjusting the start post to the preceding mile marker, while a positive end offset requires adjusting the end post to the subsequent mile marker. These adjustments ensure that the segment boundaries are accurately defined, reflecting any offsets specified in the segment data.

\begin{algorithm}[h]
\caption{Compute Seconds on Road Segment for Vehicle on Day}
\label{ag3}
\begin{algorithmic}[0]
\Function{ComputeSecondsOnRoadSegmentForVehicleOnDay}{seg, vehicleid, day}
    \If{Route Reference not present in seg}
        \State log("No Route Ref")
        \State \Return 0
    \EndIf
    \State seg\_road $\gets$ roadNameToType(seg.RouteRef)
    \State Initialize startpost, endpost, startoffset, endoffset
    
    \If{Start Post, End Post, Start Offset, and End Offset exist in seg}
        \State milemarkers $\gets$ fetchMileMarkersForRoad(road)
        \If{milemarkers is empty}
            \State \Return 0
        \EndIf
        \State startpost $\gets$ find(milemarkers, seg.StartPost)
        \State endpost $\gets$ find(milemarkers, seg.EndPost)
        \If{startpost or endpost is empty}
            \State \Return 0
        \EndIf
        \State Adjust startpost and endpost for offsets
    \EndIf
    
    \State dt $\gets$ daytracks[day][vehicleid]
    \If{dt is empty}
        \State \Return 0
    \EndIf
    \State computedSeconds $\gets$ 0
    
    \For{each (index, point) in dt.track}
        \If{point.road is empty or point.road $\neq$ seg\_road}
            \State \textbf{continue}
        \EndIf
        \If{startpost and endpost are defined}
            \State Validate point against startpost and endpost
        \EndIf

        \State next\_point $\gets$ dt.track[index + 1]
        \State duration $\gets$ min(next\_point.time.unix() - point.time.unix(), 600)
        \State computedSeconds $\gets$ computedSeconds + duration
    \EndFor
    \State computedHrs $\gets$ computedSeconds / 3600
    \State \Return computedHrs
\EndFunction
\end{algorithmic}
\end{algorithm}

The algorithm then iterates through each recorded GPS point in the vehicle's track. These GPS points already have the nearest identifiable road precomputed using geohashing. For each point, it then verifies whether the point lies on the specified road segment. If the segment has defined start and end posts, it further validates the point's position relative to these posts and their offsets. If the point satisfies all these conditions, the algorithm calculates the duration to the next point by computing the time difference between the current and next points. This duration is capped at 10 minutes to handle potential long gaps in the data. The hard threshold of 10 minutes is chosen based on sampling statistics of GPS points in the dataset where 14\% points have gaps that extend beyond 5 minutes. This helps in ensuring that any prolonged discontinuities do not skew the results. The calculated duration is then aggregated and returned as the final result as well as exported as the application's output depending on the choice of verification and creation. An example of a verification case is shown in \textbf{Figure \ref{sample_segment_case}}, where the value is displayed in the summary table as well as the output reports in \textbf{Figure \ref{sample_verify_report}} and \textbf{\ref{sample_created_records}}.

\section{Discussion}

In this section, we discuss the real-world application of our system, providing examples that highlight its potential. 

\subsection{Time Granularity in Work Orders}

A significant challenge in the telematics dataset is the lack of granularity in the start and end times for work orders (WOs). This limitation complicates the process of obtaining exact verification matches between the telematics data and the work orders. As shown in \textbf{Table \ref{tab:wo_distribution}}, the majority of work orders (21,655) in our dataset are confined to a single day. However, there are notable instances where work orders extend across multiple days.

\begin{table}[h]
    \centering
    \begin{tabular}{|c|c|}
    \hline
    \textbf{Days Spread} & \textbf{Work Order Count} \\ \hline
    1 day & 21,655 \\ \hline
    2 days & 989 \\ \hline
    3 days & 47 \\ \hline
    4 days & 2 \\ \hline
    5 days & 4 \\ \hline
    6 days & 0 \\ \hline
    7 days & 2 \\ \hline
    8 days & 1 \\ \hline
    \end{tabular}
    \caption{Distribution of Work Orders Across Multiple Days.}
    \label{tab:wo_distribution}
\end{table}

The key issue is that each work order entry in our dataset has only one date, without a precise timestamp. For instance, one work order spans eight days, involving more than ten different vehicles, highlighting the challenges due to the absence of precise start and end times. Ideally, we need two timestamps (start and end) for each work order to enhance accuracy and clarity. This lack of temporal granularity leads to several implications:

\begin{itemize}
    \item \textbf{Verification Difficulty:} Human-entered work orders may not use the same midnight-to-midnight convention as the algorithms, leading to difficult-to-detect errors in work order verification.
    \item \textbf{Extended Work Orders:} Work orders may extend beyond the recorded date, either starting earlier or ending later, leading to potential inaccuracies.
    \item \textbf{Data Entry Errors:} Manual entry can result in work orders spanning non-consecutive days or involving multiple vehicles.
\end{itemize}

Integrating our system into the work order management process can effectively address these issues. Our automated system is capable of incorporating precise start and end timestamps when creating work records, significantly enhancing the accuracy and verification of work order records. 

\subsection{Calculation of Work Hours for Drivers}

During storm events, drivers often work extended hours, and manually logging these hours becomes both a distraction and a significant burden. The manual process of recording work orders is time-consuming and prone to errors and delays. In severe snowstorms, the high volume of work orders can overwhelm manual systems, causing delays in processing drivers' salaries and resulting in dissatisfaction among drivers who rely on timely payment for their extended shifts. From a managerial perspective, manually creating work records is necessary but low-value, diverting time from strategic activities such as planning and resource allocation.

Our application simplifies this process by automating the recording of work hours. Instead of manually logging each entry, our system calculates the total work hours by summing the operational time of the given vehicle ID for the given day. This approach streamlines the process, reduces errors, and ensures accurate and timely reporting of work hours.  In order to be able to translate vehicle total work hours into driver work hours, we will need a means to identify which drivers are driving which vehicles.  We currently do not have access to such a dataset, however an exported spreadsheet of vehicle drive times are easily annotated with which driver was in which vehicle for the cases where the same driver was in the same vehicle for the entire day.

\subsection{Visualization of Overlapping Roads When Tracking Vehicles}

Tracking vehicle movements accurately, especially during winter maintenance operations, often involves dealing with overlapping road segments. Vehicles may travel the same routes multiple times, making it challenging to distinguish between individual trips and visualize the exact paths taken. This complexity can lead to difficulties in analyzing data, understanding operational patterns, and ensuring accurate reporting.

Our application addresses these challenges with advanced visualization techniques designed to depict overlapping roads. The application traces the path of each vehicle using distinct temporally color-coded lines and user-driven animation via a time slider to represent different trips on overlapping road segments. This allows users to intuitively differentiate between multiple passes over the same road.

\section{Future Work}

One notable challenge in our dataset is the presence of overlapping road segments in the total set of known road segments from INDOT.  This complicates any cumulative driving time calculations based on segment-level drive times since the same underlying GPS points will be present in multiple segments.  To address this issue, we intend to create a synthetic Road Segment list for every known road in which each segment is simply the 1-mile section between each mile marker.  Cumulative times can then be computed from this sub-set of complete but non-overlapping segments.

Future access to additional datasets such as snowplow telematics data about plow raise/lower positions and de-icing distribution equipment status will enable more accurate snow plow activity identification and segment-level de-icing product quantities.  Also, future access to driver login records to telematics systems in the snowplow trucks themselves would enable the full creation of driver-based work orders from the daily drive time totals.

\section{Conclusion}

Our paper presents a comprehensive solution to improve the work recording process for winter road maintenance operations managed by the \gls{indot}. We developed an in-browser web application that automates the creation and verification of work orders using GPS data from telematics systems. This application minimizes the need for manual data entry, enhances the granularity and accuracy of work records, and supports better resource allocation and planning.  Due to its in-browser nature, it is easily available for INDOT use without complex IT infrastructure or privacy concerns as data never leaves the user's computer.  We also outline future directions, focusing on developing advanced algorithms to detect and adjust for overlapping road segments, further refining the accuracy of total driving time calculations.

In conclusion, integrating telematics data with automated work order systems provides significant potential to advance winter road maintenance management. Our web application serves as a robust tool for INDOT and similar organizations, ensuring efficient, accurate, and reliable fleet operations.

\section{Acknowledgements}
This work was supported by the Joint Transportation Research Program (JTRP) administered by the \acrlong{indot} and Purdue University under SPR-4605. The contents of this paper reflect the views of the authors, who are responsible for the facts and the accuracy of the data presented herein. The contents do not necessarily reflect the official views and policies of the INDOT or the Federal Highway Administration. The paper does not constitute a standard, specification, or regulation.

\bibliographystyle{TRR}
\bibliography{main}

\begin{thebibliography}{10}
\providecommand{\url}[1]{\texttt{#1}}
\providecommand{\urlprefix}{URL }
\expandafter\ifx\csname urlstyle\endcsname\relax
  \providecommand{\doi}[1]{doi:\discretionary{}{}{}#1}\else
  \providecommand{\doi}{doi:\discretionary{}{}{}\begingroup \urlstyle{rm}\Url}\fi
\providecommand{\eprint}[2][]{\url{#2}}

\bibitem{FederalHighwayAdministration2023}
{Federal Highway Administration}.
\newblock Snow and ice, 2023.
\newblock \urlprefix\url{https://ops.fhwa.dot.gov/weather/weather_events/snow_ice.htm}.
\newblock Accessed: Jan. 10, 2024 [Online].

\bibitem{Zhang2024}
Zhang, Y., A.~Ault, and J.~Krogmeier.
\newblock Automated Record Keeping for Statewide Winter Road Maintenance using Telematics Tracks.
\newblock In \emph{2024 IEEE 99th Vehicular Technology Conference (VTC2024-Spring)}. Singapore, 2024.
\newblock To appear.

\bibitem{IndianaDepartmentOfTransportation}
{Indiana Department of Transportation}.
\newblock Winter operations, 2024.
\newblock \urlprefix\url{https://www.in.gov/indot/safety/winter-driving-safety-tips/winter-operations/}.
\newblock Accessed: Jan. 10, 2024 [Online].

\bibitem{IndianaDepartmentOfTransportation2023}
{Indiana Department of Transportation}.
\newblock Division of Maintenance, INDOT Work Performance Standards, 2024.
\newblock \urlprefix\url{https://www.in.gov/indot/div/pubs/INDOT-Work-Performance-Standards.pdf}.
\newblock Accessed: Jan. 10, 2024 [Online].

\bibitem{Ghaffarpasand2022}
Ghaffarpasand, O., M.~Burke, L.~K. Osei, H.~Ursell, S.~Chapman, and F.~D. Pope.
\newblock Vehicle Telematics for Safer, Cleaner and More Sustainable Urban Transport: A Review.
\newblock \emph{Sustainability (Switzerland)}, Vol.~14.
\newblock \doi{10.3390/su142416386}.

\bibitem{Wang2021}
Wang, L., M.~Ciliberto, H.~Gjoreski, P.~Lago, K.~Murao, T.~Okita, and D.~Roggen.
\newblock Locomotion and Transportation Mode Recognition from GPS and Radio Signals: Summary of SHL Challenge 2021.
\newblock In \emph{UbiComp/ISWC 2021 - Adjunct Proceedings of the 2021 ACM International Joint Conference on Pervasive and Ubiquitous Computing and Proceedings of the 2021 ACM International Symposium on Wearable Computers}. 2021.
\newblock \doi{10.1145/3460418.3479373}.

\bibitem{Zhang2020}
Zhang, Y., J.~V. Krogmeier, A.~Ault, and D.~Buckmaster.
\newblock APT3: Automated product traceability trees generated from GPS tracks.
\newblock \emph{Transactions of the ASABE}, Vol.~63.
\newblock \doi{10.13031/TRANS.13384}.

\bibitem{Kinable2016}
Kinable, J., W.~J. van Hoeve, and S.~F. Smith.
\newblock Optimization models for a real-world snow plow routing problem.
\newblock In \emph{Lecture Notes in Computer Science (including subseries Lecture Notes in Artificial Intelligence and Lecture Notes in Bioinformatics)}, Vol. 9676. 2016.
\newblock \doi{10.1007/978-3-319-33954-2_17}.

\bibitem{Goel2008}
Goel, A.
\newblock Fleet telematics: Real-time management and planning of commercial vehicle operations.
\newblock \emph{Operations Research/ Computer Science Interfaces Series}, Vol.~40.

\bibitem{Zhang2017}
Zhang, Y., A.~Ault, J.~V. Krogmeier, and D.~Buckmaster.
\newblock Activity Recognition for Harvesting via GPS Tracks.
\newblock In \emph{2017 ASABE Annual International Meeting}. American Society of Agricultural and Biological Engineers, 2017, p.~1.
\newblock \doi{10.13031/aim.201700813}.

\bibitem{Niemeyer2008}
Niemeyer, G.
\newblock Geohash.
\newblock \emph{Retrieved June}, Vol.~6.

\end{thebibliography}
\end{document}